\documentclass[12pt]{article}

\newcommand{\df}{\partial}
\newcommand{\beq}{\begin{equation}}
\newcommand{\eeq}{\end{equation}}
\newcommand{\beqn}{\begin{eqnarray}}
\newcommand{\eeqn}{\end{eqnarray}}

\title{The Off-Shell Boundary State and Cross-Caps in the Genus Expansion of String Theory}

\author{M. Laidlaw \\Department of Physics and Astronomy\\
University of British Columbia\\Vancouver, British Columbia, Canada V6T 1Z1}

\begin{document}
\maketitle

\abstract{In this paper we use the boundary state formalism for the bosonic string to calculate the 
emission amplitude for closed string states from particular D-branes.  We show that the amplitudes 
are exactly those obtained from world-sheet sigma model calculations, and the construction of the 
boundary state automatically enforces the requirement for integrated vertex operators, even in the 
case of an off-shell boundary state.  Using the boundary state and a similar expansion for the 
cross-cap, we  produce higher order terms in the string loop expansion for the partition 
function of the quadratic backgrounds considered.}

\newpage

\section{Introduction}

The study of off shell string theory has been addressed many times in the literature
 within the context 
of
background independent string
field theory
\cite{Witten:1992qy,Witten:1993cr,Witten:1993ed,Shatashvili:1993kk,Shatashvili:1993ps}
which 
has been the subject of a considerable amount of interest in that it can provide useful 
information about the properties of unstable d-branes \cite{Gerasimov:2000zp,
Gerasimov:2000ga,Kutasov:2000qp}.
  Dispite this
there are several subtleties that have been examined, and in particular  
a great deal of effort has been 
expended
in determining an action for a tachyon field coupled to a bosonic string \cite{Gerasimov:2000zp,
Gerasimov:2000ga,Kutasov:2000qp,Akhmedov:2001jq,Gerasimov:2001pg,Kraus:2000nj,
Craps:2001jp,
Viswanathan:2001cs,Rashkov:2001pu,Alishahiha:2001tg,Andreev:2000yn,Arutyunov:2001nz,
deAlwis:2001hi}
, and while 
great progress had been made the understanding of higher loop effects is incomplete at best. 

A tractable problem 
within this genre is the study of the off-shell theory in the background of a quadratic tachyon 
potential, a problem that is similar in spirit and detail to the examination of string theory in the 
background of a constant electromagnetic field. \cite{Fradkin:1985qd}
In this paper we combine these naturally compatible studies using the boundary state formalism
 \cite{Bardakci:2001ck,
Lee:2001cs,Lee:2001ey,Fujii:2001qp,DiVecchia:1999fx,Akhmedov:2001yh}.  It 
allows us to calculate the probability for a topological defect which supports these quadratic 
fields to emit any number of closed string states into its bulk space-time.  The loss of conformal 
invariance introduced by the background tachyon field is naturally accomodated by a conformal 
transformation which induces a calculable change in the boundary state.  This new boundary state can 
be shown to reproduce the sigma model expectation values for the insertion of a vertex operator at 
an arbitrary point on the string world-sheet.  

Using the correspondence between the sigma model calculation and that in the operator formalism the 
question of higher genus surfaces with some number of boundaries interacting with the background 
fields is considered.  The insertion of both loops and boundaries is included naturally in this 
method, and the results obtained are compared with known results.

Throughout this work the action under consideration is
\beqn
S\left( g, F, T_0, U \right) &=&
\frac{1}{ 4 \pi \alpha'} \int_\Sigma d\sigma d\phi ~ g_{\mu\nu}  \df^a X^\mu \df_a X_\mu 
\nonumber \\
&~& + \int_{\df \Sigma} d\phi \left(\frac{1}{2} F_{\mu\nu} X^\nu \df_\phi X^\mu + \frac{1}{2\pi}
T_0 + \frac{1}{8\pi} U_{\mu\nu} X^\mu X^\nu \right),
\label{action}
\eeqn
where $\alpha'$ is the inverse string tension, $\Sigma$ is the string world-sheet, $d\sigma d\phi$ 
is the integration measure of the string bulk, $d\phi$ is the integration measure of the string 
world-sheet boundary, and $\df_\phi$ is the derivative tangential to that boundary.  The field 
content in this are a constant $U(1)$ gauge field $F_{\mu\nu}$ 
and the tachyon profile $T(X) = \frac{1}{2\pi}
T_0 + \frac{1}{8\pi} U_{\mu\nu} X^\mu X^\nu$ is characterized by a constant, $T_0$ and a constant 
symmetric matrix $U_{\mu\nu}$ which provides a simple generalization for the discussion given in 
{\cite{Craps:2001jp}}.

\section{The Boundary State}

The virtue of the boundary state as a tool in the analysis of the action above is that it 
allows calculations that previously took careful integration to be reduced to algebraic 
manipulations.  We wish to carefully construct the boundary state and to show that it 
reproduces with ease the particle emission amplitudes that would be obtained from the string 
sigma model.  The starting point for this analysis is the action (\ref{action}).  By varying 
it, 
one obtaines the equation
\beq
\left( \frac{1}{2 \pi \alpha'} g_{\mu\nu} \df_\sigma 
 + F_{\mu\nu} \df_\phi  + \frac{1}{4\pi} U_{\mu\nu} \right) X^\nu = 0
\label{bosonicBCfields}
\eeq
as the boundary condition for the string world-sheet.  Recalling the conventions from the 
action, $\df_\sigma$ is the derivative normal to the boundary and $\df_\phi$ is the tangential 
to the boundary.  We now create a state $| B \rangle$ that obeys the above condition as an 
operator equation.  To do this we use reparametrize the string world sheet in terms of 
holomorphic and antiholomorphic variables $z=\sigma e^{i\phi}$ and $\bar z = \sigma e^{-i\phi}$
and use the standard mode expansion for $X$ as a function of $z$
\beq
X^\mu(z,\bar z) = x^\mu + p^\mu \ln |z^2| + \sum_{m \neq 0} \frac{1}{m} \left(
\frac{\alpha_m^\mu}{z^m} + \frac{ \tilde \alpha_m^\mu}{\bar z^m} \right).
\eeq
we find that in terms of the mode operators the boundary conditions read
\beq
\left( g + 2\pi \alpha' F + \frac{\alpha'}{2} \frac{U}{n} \right)_{\mu\nu} \alpha^\mu_n
+ \left( g - 2\pi \alpha' F - \frac{\alpha'}{2} \frac{U}{n} \right)_{\mu\nu} \tilde
\alpha^\mu_{-n} = 0.
\label{bosonicBC}
\eeq 
The condition for the boundary state to obey (\ref{bosonicBCfields})
can then be restated in terms of (\ref{bosonicBC})
to be 
\beq 
\left[\left( g + 2\pi \alpha' F + \frac{\alpha'}{2} \frac{U}{n} \right)_{\mu\nu}
\alpha^\mu_n
+ \left( g - 2\pi \alpha' F - \frac{\alpha'}{2} \frac{U}{n} \right)_{\mu\nu} \tilde
\alpha^\mu_{-n} \right] |B\rangle = 0,
\nonumber \\
\eeq
\beq
\left[ g_{\mu\nu} p^\mu - i \frac{\alpha'}{2} U_{\mu\nu} x^\mu \right] |B\rangle
 = 0.
\label{conditiononP}
\eeq
To satisfy this it is clear that $|B \rangle$ must be a coherent state, and it is given by
\cite{Akhmedov:2001yh}
\beqn
|B\rangle &=& {\cal N} \prod_{n\geq1} \exp  \left( -
\left( \frac{g - 2 \pi \alpha' F - \frac{\alpha'}{2} \frac{U}{n} }{
 g + 2 \pi \alpha' F + \frac{\alpha'}{2} \frac{U}{n} } \right)_{\mu\nu}
\frac{\alpha^\mu_{-n} \tilde \alpha^\nu_{-n} }{n} \right)
\exp \left(- \frac{\alpha'}{4} x^\mu U_{\mu\nu} x^\nu \right) | 0 \rangle
\nonumber \\
&=&  {\cal N} \prod_{n\geq1} 
\exp  \left( - \Lambda^n_{\mu\nu} \alpha^\mu_{-n} \tilde \alpha^\nu_{-n} \right) 
\exp \left(- \frac{\alpha'}{4} x^\mu U_{\mu\nu} x^\nu \right) | 0 \rangle
\eeqn
where ${\cal N}$ is a normalization constant which must be determined, and we define
\beq
\Lambda^{n}_{\mu\nu} = \frac{1}{n} 
\frac{g - 2 \pi \alpha' F - \frac{\alpha'}{2} \frac{U}{n} }{
 g + 2 \pi \alpha' F + \frac{\alpha'}{2} \frac{U}{n} }
\eeq
for future convenience.

\subsection{Conformal Transformation of the Boundary State}

Clearly this boundary state is not conformally invariant due to the addition of the interaction 
with the tachyon field.  The two cases where we expect conformal invariance are at the two 
fixed points of renormalization group flow, namely $U=0$ and $U=\infty$
which correspond respectively to the case of Neumann or Dirichlet boundary conditions on the 
boundary of the string world sheet.  Note that in the case of Dirichlet boundary conditions the 
interaction with the background electromagnetic field is eliminated, as would be expected from 
the sigma model point of view.  
Because of this it is interesting to examine how the boundary state transforms under the 
PSL(2,R) symmetry that is broken by the presence of the $U$ term in the boundary state.  In the 
two conformally invariant cases this leaves the action invariant.
The action of PSL(2,R) on the complex coordinates of the
disk is to perform the mapping
\beq
z \rightarrow w(z) = \frac{a z + b}{b^* z + a^*}
\label{conftranfandnorm}
\eeq
where $a$ and $b$ satisfy the relation
\beq
|a^2| - |b^2| = 1.
\eeq
This transformation maps the interior of the unit disk to itself, the exterior
to the exterior and the boundary to the boundary.
Moreover, this transformation of the coordinates induces a
mapping which intermixes the oscillator modes.
To see this consider the definition of the oscillator modes
To see this consider the definition of the oscillator modes
\beq
\alpha_m^\mu = \sqrt{ \frac{2}{\alpha'} }
\oint \frac{dz}{2\pi } z^m \df X^\mu(z)
\eeq
where the contour is the boundary of the unit disk, and the mode expansion of 
$X$ is
\beq
\df X^\mu(z) = -i \sqrt{ \frac{\alpha'}{2} } \sum_m \frac{\alpha_m^\mu}{z^{m+1} }.
\label{modeexp}
\eeq
Now, using the fact that $X$ is a scalar, or equivalently the fact that
$\df X$ is a (1,0) tensor, we see that
\beq
\alpha_m^\mu = \oint \frac{dz}{2\pi i} z^m \df_w X^\mu(w) \frac{dw}{dz}
.
\eeq
Now, using the fact that a mode expansion for $X$ exists in terms of $w$ with 
coefficients $\alpha'_m$ in exactly the same way as (\ref{modeexp}), we see that
\beq
\alpha_m^\mu = M^{(a,b)}_{mn} \alpha_n^{'\mu}
\eeq
where
\beq
M^{(a,b)}_{mn} = \oint \frac{dz}{2\pi i} z^m \frac{ (b^* z + a^*)^{n-1} }{
(a z + b)^{n+1} }.
\label{defofM}
\eeq  

The properties of the matrix $M$ are interesting and facilitate further study.  The matrix has 
a block diagonal form so that creation and annihilation operators are not mixed by the 
conformal transformation, and with appropriate normalization of the oscillator modes it can be 
seen to be hermitian, or equivalently that it preserves the inner product on the space of 
operators.  The exact form $M$ as a function of its indices can be easily obtained, 
but for the purposes of this discussion it is easier to to simply note that with the rescaling
${\cal M}_{mp} = \sqrt{\frac{p}{m} } M_{mp}$ for either $m,p > 0$ or $m,p < 0$ then
${\cal M}^{-1}_{mp} = {\cal M}_{mp}^\dagger$.

Using this information we obtain that the modifictation of the boundary state 
\beqn
|B_{a,b} \rangle &=& {\cal N} \exp \left( \sum_{n=1, j,k=-\infty}^\infty
\alpha^{\mu}_{-k} M^{(a,b)}_{-n-k} \Lambda^n_{\mu\nu} {M^{(a,b)*}_{-n-j} } 
\tilde \alpha^{\nu}_{-j}  \right) 
\nonumber \\ &~&
\exp \left(- \frac{\alpha'}{4} x^\mu U_{\mu\nu} x^\nu \right)
 | 0 \rangle.
\label{transfboundstate}
\eeqn
In this equation and all following ones we drop the $'$ associated with the transformed 
oscillators for notational simplicity.  Due to the intuition from the conformally invariant 
cases we conjecture that the  proper definition of the boundary state to give
the correct overlap with all closed string states is
\beq
|B\rangle = \int d^2 a d^2 b \delta(|a^2| - |b^2| -1) |B_{a,b} \rangle.
\eeq
This is just the boundary state
(\ref{transfboundstate}) integrated over the Haar measure of  PSL(2,R).

\subsection{Boundary State emission of one particle}

Since we wish to show that the boundary state is an algebraized version of the action (\ref{action}) 
we must calculate the emission probability for various particles from the boundary state above.
This has been done in more detail in {\cite{Akhmedov:2001yh}} but we recapitulate the results here for 
completeness.

The case of the tachyon is straightforward.  We must evaluate the overlap of the Fock space ground 
state with the  transformed boundary state (\ref{transfboundstate}).  
Here, and in subsequent formulae we omit the momentum conserving $\delta$-functions, and the 
integration over the transformation parameters for the boundary state.
For a tachyon with momentum $p^\mu$ we find that the probability for emission from the boundary 
state is
\beqn
\langle 0, p^\mu | B_{a,b} \rangle &=& {\cal N} \exp \left( - p^\mu p^\nu \frac{\alpha'}{2}
\sum_{n=1}^\infty M^{(a,b)}_{-n 0} \Lambda^n_{\mu\nu} M^{(a,b)*}_{-n 0} \right)
\nonumber \\
&=& 
 {\cal N}
 \exp \left( - p^\mu p^\nu \frac{\alpha'}{2}
\sum_{n=1}^\infty \frac{1}{n} \left( \frac{ g - 
2\pi \alpha' F - \frac{\alpha'}{2} \frac{U}{n} }
{ g+2\pi \alpha' F+\frac{\alpha'}{2} \frac{U}{n} } 
\right)_{\mu\nu} \frac{ |b|^{2n} }{ |a|^{2n} } 
\right).
\label{1tachyonemissioncalc}\eeqn
In the above expression we have used the previously defined form for $\Lambda^n_{\mu\nu}$,
the fact that $ M^{(a,b)}_{-n 0} = \left( \frac{ -b^*}{a^*} \right)^n$, and the relation 
$\alpha^\mu_o = \sqrt{\frac{\alpha'}{2} } p^\mu$.

Similarly, for an arbitrary massless state  with polarization tensor $P_{\mu\nu}$
and momentum $p^\mu$
\beqn
| P_{\mu\nu} \rangle &=&  P_{\mu\nu} \alpha^\mu_{-1}
\tilde \alpha^\nu_{-1} | 0, p^\mu  \rangle
\label{masslessdefn}
\eeqn
 the overlap
to be calculated is
\beqn
\langle   P_{\mu\nu} | B_{a,b} \rangle
&=&
{\cal N}
\exp
\left( - p_\mu p_\nu \frac{\alpha'}{2} \sum_{n=1}^\infty M^{(a,b)}_{-n0} 
\Lambda^n_{\mu\nu}
M^{(a,b)*}_{-n0} \right)
\nonumber \\
&~& P^{\mu\nu}  \left[ - \sum_{n=1}^\infty   M^{(a,b)}_{-n-1} \Lambda^n_{\mu\nu}
M^{(a,b)*}_{-n-1}   \right.
\nonumber \\
&~& +p^\alpha p^\beta \frac{\alpha' }{2} \sum_{n=1}^\infty   M^{(a,b)}_{-n-1}
\Lambda^n_{\mu\alpha} 
M^{(a,b)*}_{-n0} 
\left. \sum_{m=1}^\infty   M^{(a,b)}_{-m0}
\Lambda^m_{\beta\nu} 
M^{(a,b)*}_{-m-1}
\right]
\nonumber \\
&=&
 {\cal N}   
 \exp \left( - p^\mu p^\nu \frac{\alpha'}{2}
\sum_{n=1}^\infty \frac{1}{n} \left( \frac{ g - 2\pi \alpha' F - \frac{\alpha'}{2} \frac{U}{n} }
{ g+2\pi \alpha' F+\frac{\alpha'}{2} \frac{U}{n} } \right)_{\mu\nu} \frac{ |b|^{2n} }{ |a|^{2n} }
\right)
\nonumber \\
&~& 
 P^{\mu\nu}  \left[ - \sum_{n=1}^\infty n
\left( \frac{g - 2 \pi \alpha' F - \frac{\alpha'}{2} \frac{U}{n}   
 }{ g + 2 \pi \alpha' F + \frac{\alpha}{2} \frac{U}{n} } \right)_{\mu\nu}
 \frac{ |b|^{2(n-1)} }{ |a|^{2(n-1) } } \frac{1}{|a^2|^2}   \right.
\nonumber \\
&~& +p^\alpha p^\beta \frac{\alpha' }{2} \sum_{n=1}^\infty
\left( \frac{g - 2 \pi \alpha' F - \frac{\alpha'}{2} \frac{U}{n} }{
 g + 2 \pi \alpha' F + \frac{\alpha}{2} \frac{U}{n} } \right)_{\mu\alpha}
\frac{ |b|^{2(n-1)} }{ |a|^{2(n-1) } } \frac{ -b^* }{ |a^2| a^* }
\nonumber \\
&~&
\times \left. \sum_{m=1}^\infty
 \left( \frac{g - 2 \pi \alpha' F - \frac{\alpha'}{2} \frac{U}{m} }{
 g + 2 \pi \alpha' F + \frac{\alpha}{2} \frac{U}{m} } \right)_{\beta\nu}
\frac{ |b|^{2(m-1)} }{ |a|^{2(m-1) } } \frac{ -b}{ |a^2| a}
\right]
\label{gravitonBScalc}
\eeqn
where again the explicit form of the matrices $M$ has been used in the last equality.

This kind of arguement can be repeated indefinitely, but we present one more such calculation which 
contains the germs of generality, which will prove useful to consider.  In particular the more 
general state $A$ with momentum $p^\mu$ is defined by
\beqn
|A_{\mu\nu\delta} \rangle &=&  
 A_{\mu\nu\delta} \alpha_{-a}^\mu \tilde \alpha_{-b}
^\nu \tilde \alpha_{-c}^\delta |0,p^\mu \rangle,
\eeqn
and its overlap with the boundary state is given by
\beqn  
\langle  A_{\mu\nu\delta}
| B \rangle
&=&
{\cal N}
\exp
\left( - p_\mu p_\nu \frac{\alpha'}{2} \sum_{n=1}^\infty M^{(a,b)}_{-n0}
\Lambda^n_{\mu\nu}
M^{(a,b)*}_{-n0} \right) \times
\nonumber \\
&~&
A^{\mu\nu\delta} \sqrt{\frac{\alpha'}{2}}
 \left[ \sum_n ab M^{(a,b)}_{-n -a}    
\Lambda_{\mu\nu}^n
M^{(a,b)*}_{-n-b}
p^\alpha \sum_m c M^{(a,b)}_{-m 0} \Lambda_{\alpha \delta}^m M^{(a,b)*}_{-m-c}  \right.
\nonumber \\
&~&  + p^\alpha \sum_n ac M^{(a,b)}_{-n -a} \Lambda_{\mu\delta}^n M^{(a,b)*}_{-n-c}
\sum_m b M^{(a,b)}_{-m 0} \Lambda_{\alpha \nu}^m M^{(a,b)*}_{-m-b}
\nonumber \\ &~&
 - p^\alpha p^\beta p^\gamma 
\frac{\alpha'}{2} \sum_n aM^{(a,b)}_{-n -a} \Lambda_{\mu\alpha}^n 
M^{(a,b)*}_{-n0}
\sum_m bM^{(a,b)}_{-m 0} \Lambda_{\beta \nu}^m M^{(a,b)*}_{-m-b}
\nonumber \\ &~&
\left. \times \sum_l c M^{(a,b)}_{-l 0} 
\Lambda_{\gamma \delta}^l M^{(a,b)*}_{-l-c} 
\right].
\label{arbstateBScalc}
\eeqn
At this point the formulae become more abstruse and do not in general convey any more information 
than can already be discerned.  It should be noted that the contractions of the various matrices 
look suspiciously like those of Green's functions, which it will transpire that they are, but to see 
this requires a simple calculation.  A special case of a more general formula proven in the next 
section shows that for $y = \frac{ az + b}{ b^* z + a^*}$ subject to $|a|^2 -|b|^2 =1$ we have that
\beq
\frac{1}{(k-1)!} \df^k z^d(y) |_{y=0} = k M^{(a,b)*}_{-d-k}.
\label{specialcase}
\eeq
Note that since the transformation from $z$ to $y$ is one-to-one the above equation makes sense.
This completes the analysis for the emission of one particle from the boundary state 
$|B_{a,b}\rangle$, however the question becomes more interesting for the emission of more than one 
particle.

\subsection{Boundary State emission of many particles}

As in the case of emission of one particle by the boundary state it is perhaps the most instructive 
to consider the case of the emission of two tachyons first, and then specialize to more complicated 
correlators.
Ordering the operators appropriately 
for radial (as opposed to anti-radial) quantization and noting that the PSL(2,R) transformation is 
not sufficient to fix the location of both close string vertex operators we proceed to calculate, 
using the previous definitions and mode expanstion
\beqn
\langle B_{a,b} | \left. : \exp\left( i k^\mu X_\mu \right) :
\right|_{\omega} | 0,p^\mu \rangle
&=& {\cal N} \langle 0 | \exp \left( - \sum \alpha_i^\mu M^{(a,b)}_{n i} \Lambda^n_{\mu\nu} 
M^{(a,b)*}_{n j} \tilde \alpha ^\nu_j \right)
\nonumber \\ &~&
\exp\left(
  k_\mu \sqrt{\frac{\alpha'}{2} } \sum_{l>0} \frac{1}{l} \left( \alpha^{\mu}_{ -l} 
\omega^l 
+ \tilde \alpha^{\mu}_{ -l}  \bar \omega^l \right)  \right)
\nonumber \\
&~& \exp \left(i k^\mu x_\nu + \sqrt{\frac{\alpha'}{2} } k_\mu \alpha^\mu_0 \ln |\omega|^2 \right)
| 0,p^\mu \rangle
\nonumber \\
&=&
{\cal N}
\exp \left( k^\mu p_\mu \frac{ \alpha'}{2} \ln |\omega|^2 \right)
\exp \left( - p^\mu p^\nu \frac{ 
\alpha'}{2} \sum_{n=1}^\infty M^{(a,b)}_{n 0} \Lambda^n_{\mu\nu} 
M^{(a,b)*}_{n 0} 
\right) 
\nonumber \\
&~&
\exp \left( - p^\mu k^\nu \frac{ \alpha'}{2} \sum_{n=1,j=0}^\infty M^{(a,b)}_{n 0} 
\Lambda^n_{\mu\nu} M^{(a,b)*}_{n j}
\bar \omega^j
\right)
\nonumber \\
&~&
\exp \left( - k^\mu p^\nu \frac{ \alpha'}{2} \sum_{n=1,i=0}^\infty \omega^i M^{(a,b)}_{n i} 
\Lambda^n_{\mu\nu} 
M^{(a,b)*}_{n 0}
\right)
\nonumber \\
&~&
\exp \left( - p^\mu k^\nu \frac{ \alpha'}{2} 
\sum_{n=1,i,j=0}^\infty \omega^i M^{(a,b)}_{n i} 
\Lambda^n_{\mu\nu} 
M^{(a,b)*}_{n j}
\bar \omega^j
\right).
\label{twotachyonamp}
\eeqn 
Upon closer inspection it is apparent that this very reminiscent of a pair of 
exponentiated greens functions.  In fact, we will show in the next section that the 
expectation value of two tachyon vertex operators in the 
background of the boundary interaction exactly coincides with this.

The next natural quantity to calculate is the emission of a more general state in place of 
either, or both tachyons in the previous calculation.  It is of course possible to demonstrate 
the overlap of an arbitrary string state explicitly, but the combinatorial nature of the 
result quickly renders the resulting expressions obscure.  With this in mind we examine a 
state that contains the germ of generality and corresponds to the calculation done in the case 
of one particle emission.  
\beqn
\langle B_{a,b} | \left. : \exp\left( i k^\mu X_\mu \right) :
\right|_{\omega} A_{\mu\nu\delta} \alpha_n^\mu 
\tilde \alpha_p^\nu \tilde \alpha_q^\delta| 0,p^\mu \rangle
= {\cal A}_{{\cal T} 2} \times A^{\mu\nu\delta} 
\nonumber \\
\left[ \sqrt{ \frac{\alpha'}{2} }^3 
\left( - \sum n M_{rn}^{(ab) } \Lambda_{\mu \gamma}^r 
M_{rj}^{(ab)*} \bar \omega^j k^\gamma - k_\mu \frac{1}{\omega^n} - \sum 
n M_{rn}^{(ab) } \Lambda_{\mu \gamma}^r
M_{r0}^{(ab)*} p^\gamma \right) \right.
\nonumber \\
\left( - \sum k^\gamma \omega^i M_{ri}^{(ab)} \Lambda^r_{\gamma \nu} M^{(ab)}*_{rp} p
- k_\nu \frac{1}{\bar \omega^p} - \sum p^\gamma M_{r0}^{(ab)}  \Lambda^r_{\gamma \nu} 
M^{(ab)*}_{rp} p
\right)
\nonumber \\
\left( - \sum k^\gamma \omega^i M_{ri}^{(ab)} \Lambda^r_{\gamma \delta} M^{(ab)*}_{rq} q
- k_\nu \frac{1}{\bar \omega^q} - \sum p^\gamma M_{r0}^{(ab)}  \Lambda^r_{\gamma \delta}
M^{(ab)*}_{rq} q
\right)
\nonumber \\
 + 
\left( - \sum k^\gamma \omega^i M_{ri}^{(ab)} \Lambda^r_{\gamma \delta} M^{(ab)*}_{rq} q
- k_\nu \frac{1}{\bar \omega^q} - \sum p^\gamma M_{r0}^{(ab)}  \Lambda^r_{\gamma \delta}
M^{(ab)*}_{rq} q
\right)
\nonumber \\
\left. \left( - \sum n M_{rn}^{(ab) } \Lambda_{\mu \nu}^r
 M^{(ab)*}_{rp} p
\right) \sqrt{ \frac{\alpha'}{2} }
 \right] + \left( p\leftrightarrow q, \nu \leftrightarrow \delta \right).
\label{twovertexbigmess2}
\eeqn
In the above $ {\cal A}_{{\cal T} 2}$ is the result for the boundary state to emit two 
tachyons, which appears as a multiplicative factor and is calculated explicitly above 
(\ref{twotachyonamp}).

Similarly it is possible to calculate the analogous expression for the vertex which emits the 
complicated state at the point $\omega $ on the disk, and using the standard commutation 
relationships as outlined previously we find
\beqn
\langle B_{a,b} | \left. : A_{\mu\nu\delta} \frac{1}{(n-1)!} \df^n X^\mu 
\frac{1}{(p-1)!} \bar \df^p X^\nu \frac{1}{(q-1)!} \bar \df^q X^\delta 
\exp\left( i k^\mu 
X_\mu \right) :
\right|_{\omega} | 0,p^\mu \rangle
=
{\cal A}_{{\cal T} 2} \times A^{\mu\nu\delta}
\nonumber \\
\left[ - \left( \sum \frac{1}{(n-1)!} \frac{1}{(p-1)!} \frac{m!}{(m-n)!} 
\omega^{m-n}
M_{rm}^{(ab)} \Lambda_{\mu\nu}^r M_{rj}^{(ab)*} \frac{j!}{(j-p)!} 
\bar 
\omega^{j-p} \right) 
\right.
\nonumber \\ 
+ \left\{ \frac{\alpha'}{2} 
\left( - \sum \frac{1}{(n-1)!}  \frac{m!}{(m-n)!} \omega^{m-n}
M_{rm}^{(ab)} \Lambda_{\mu \gamma}^r M^{(ab)*}_{rj} \bar \omega^j k^\gamma  \right. \right. 
\nonumber \\
\left.
- \sum \frac{1}{(n-1)!}  \frac{m!}{(m-n)!} \omega^{m-n}
M_{rm}^{(ab)} \Lambda_{\mu \gamma}^r M^{(ab)*}_{r0} p^\gamma + p_\mu (-1)^n \omega^{-n} 
\right)
\nonumber \\
\left( - \sum p^\gamma M^{(ab)}_{r0} \Lambda^r_{\gamma \nu} M^{(ab)*}_{rj} \frac{1}{(p-1)!}
\frac{j!}{(j-p)!}\bar \omega^{j-p}
\right. \nonumber \\
\left. \left.
\left. -  \sum k^\gamma \omega^m M^{(ab)}_{rm} \Lambda^r_{\gamma \nu} M^{(ab)*}_{rj} 
\frac{1}{(p-1)!}
\frac{j!}{(j-p)!}\bar \omega^{j-p} +  p_\mu (-1)^p \bar \omega^{-p} \right) \right\} \right]
\nonumber \\
\left( - \sum p^\gamma M^{(ab)}_{r0} \Lambda^r_{\gamma \delta} M^{(ab)*}_{rj} 
\frac{1}{(q-1)!}
\frac{j!}{(j-q)!}\bar \omega^{j-q}        
\right. \nonumber \\
\left. -  \sum k^\gamma \omega^m M^{(ab)}_{rm} \Lambda^r_{\gamma \delta} M^{(ab)*}_{rj}
\frac{1}{(q-1)!}
\frac{j!}{(j-q)!}\bar \omega^{j-q} +  p_\mu (-1)^q \bar \omega^{-q} \right)
 + \left( p\leftrightarrow q, \nu \leftrightarrow \delta \right).
\label{twovertexbigmess}
\eeqn
The above expression can be seen to be the same as that of the emission with the complicated 
vertex at the center, as the case of two tachyon emission would suggest.

\section{Sigma Model}
Having performed an the calculations from the point of view of the raising and lowering 
operators it is now instructive to compare with what should be analogous results from sigma 
model calculations.
We fix our convention that the functional integral is in all cases the
average over the action given in (\ref{action}),
\beq
\langle {\cal O}(X) \rangle = \int {\cal D} X e^{-S(X)}  {\cal O}(X).
\eeq
In addition, the greens function on the unit disk with Neumann boundary conditions
is determined to be \cite{Hsue:1970ra} 
\beq
G^{\mu\nu}(z,z') = - \alpha' g^{\mu\nu} \left( -  \ln \left| z-z' \right| - \ln
\left| 1- z \bar z' \right| \right),
\label{diskprop}  
\eeq
and it will be useful also to know the bulk to boundary propagator
which is
\beq
G^{\mu\nu}(\rho e^{i \phi}, e^{i \phi'} ) = 2 \alpha' g^{\mu\nu}
\sum_{m=1}^\infty \frac{\rho^m}{m}
\cos[ m(\phi - \phi') ].
\label{diskbbprop}
\eeq
The boundary to boundary propagator can be read off from (\ref{diskbbprop}) as
the limit in which $\rho \rightarrow 1$.
Throughout, we will use $z=\rho e^{i \phi}$ as a parameterization of the points within
the unit disk, so $0 \leq \rho \leq 1$ and $0 \leq \phi <  2\pi$.
Using the bulk to boundary propagator it is possible to integrate out the quadratic
interactions on the boundary {\cite{Fradkin:1985qd}} and to obtain an exact propagator, 
which is given by
\beqn
G^{\mu\nu} (z,z') &=&
-\alpha' g^{\mu\nu} \ln \left| z - z' \right|
+ \frac{\alpha'}{2}
\sum_{n=1}^\infty \left( \frac{g - 2 \pi \alpha' F - \frac{ \alpha'}{2}
\frac{U}{n} }{ g + 2 \pi \alpha' F + \frac{ \alpha'}{2} \frac{U}{n} } \right)^{ 
\left\{ \mu\nu\right\} }
\frac{(z \bar z')^n + (\bar z z')^n}{n}
\nonumber \\
&~& +
\alpha' \sum_{n=1}^\infty \left( \frac{  2 \pi \alpha' F + 
\frac{ \alpha'}{2}
\frac{U}{n} }{ g + 2 \pi \alpha' F + \frac{ \alpha'}{2} \frac{U}{n} } \right)^{
\left[\mu\nu \right]}
\frac{(z \bar z')^n - (\bar z z')^n}{n}
\nonumber \\
&=&
-\alpha' g^{\mu\nu} \ln \left| z - z' \right|
+ \frac{\alpha'}{2} 
\sum_{n=1}^\infty \left( \frac{g - 2 \pi \alpha' F - \frac{ \alpha'}{2}
\frac{U}{n} }{ g + 2 \pi \alpha' F + \frac{ \alpha'}{2} \frac{U}{n} } \right)^{ 
\left\{ \mu\nu \right\} }
\frac{(z \bar z')^n + (\bar z z')^n}{n}
\nonumber \\
&~& +
\frac{\alpha'}{2} \sum_{n=1}^\infty \left( \frac{ g - 2 \pi \alpha' F -
\frac{ \alpha'}{2}
\frac{U}{n} }{ g + 2 \pi \alpha' F + \frac{ \alpha'}{2} \frac{U}{n} } \right)^{
\left[\mu\nu \right] }
\frac{(z \bar z')^n - (\bar z z')^n}{n}.
\label{bosonicgf}
\eeqn
Note that this expression is symmetric as it should be because the antisymmetry of 
lorentz indices in the final term is compensated by the antisymmetry of the coordiate 
term.

The first calculation that must be done to determine the normalization of the 
sigma model amplitudes is the partition function.  In this approach the 
oscillator modes of $X$ must be integrated out with the contributions from $F$ and $U$
treated as perturbations.  Since both perturbations are quadratic, all the feynmann 
graphs
that contribute to the free energy can be written and evaluated, and explicitly the 
free energy is given by 
\beq
{\cal F} =
 - \sum_{m=1}^\infty Tr \ln{ \left( g + 2\pi \alpha' F + \frac{\alpha'}{2} \frac{U}{m}  
\right)  },
\label{messyeqn}
\eeq
{ see \cite{Fradkin:1985qd,Laidlaw:2000kb} for further calculations done in this 
spirit. }
From
(\ref{messyeqn})
we immediately see that the partition function is given by   
\beqn
Z &=& e^{-T_0} \prod_{m=1}^\infty \frac{1}{\det \left(
g + 2\pi \alpha' F + \frac{\alpha'}{2} \frac{U}{m} \right) }
\int dx_0 e^{- \frac{ U_{\mu\nu} }{4} x_0^\mu x_0^\nu }
\nonumber \\   
&=& \frac{1 }{\det \left( \frac{U}{2}  \right) }  e^{-T_0} \prod_{m=1}^\infty 
\frac{1}{\det \left(
g + 2\pi \alpha' F + \frac{\alpha'}{2} \frac{U}{m} \right) }.
\label{diskpf}
\eeqn
This expression is divergent, but using $\zeta$-function regularization   
{\cite{Kraus:2000nj} } it
can be reduced to
\beq
Z = e^{-T_0} \sqrt{ \det\left( \frac{g + 2\pi \alpha' F}{U/2 } \right) }
\det \Gamma \left( 1 + \frac{\alpha' U /2}{g + 2\pi \alpha' F} \right),
\label{partitionfunction}
\eeq
where $\Gamma(g)$ is the $\Gamma$ function and the dependence of all transcendental 
functions
on the matrices $U$ and $F$ is defined by their Taylor expansion.

\subsection{Conformal transformation on the disk}

We now wish to calculate the expectation value for vertex operators that correspond to 
different
closed string states, however this is a process that must be done with some care.
To  calculate the emission of a closed string in the world-sheet picture
one generally  considers
 a disk emitting an asymptotic closed string state.  This is really a closed string
cylinder
diagram. The standard method is to use
 conformal invariance to map the closed string state to a point
on
the disk, namely the origin, where a corresponding vertex operator is inserted.
On the other hand it has been cogently
argued that it is necessary to have an integrated vertex operator for closed string 
states to
properly couple
{\cite{Craps:2001jp}}, in particular that the graviton must be
produced by an integrated vertex operator to
couple correctly to the energy momentum tensor.
The distinction between a fixed vertex operator and an integrated vertex
operator is moot in the conformally invariant case where the integration
will only produce a trivial volume factor, however in the case we consider
more care must be taken.  We wish to consider arbitrary locations of the
vertex operators on the string world sheet, and the natural measure to
impose is that of the conformal transformations which map the origin to
a point within the unit disk on the complex plane.

In other words we propose to allow the vertex
operator corresponding to the closed string state to be moved from the origin by a 
conformal
transformation that preserves the area of the unit disk, namely a PSL(2,R) 
transformation.
The method to accomplish this is to go to a new coordinate system
\beq
y = \frac{ a z + b }{b^* z + a^*},~~|a^2| - |b^2| = 1,
\eeq
and a vertex operator at the origin $y=0$ would correspond to an insertion of a vertex 
operator
at the point $z = \frac{-b}{a}$.  It is worth noting that in the case of conformal 
invariance,
that is when $U \rightarrow 0$ or $U \rightarrow \infty$ the greens function remains 
unchanged
in form, the $y$
dependence coming  from  the replacement $z \rightarrow z(y)$. Even in the case of finite 
$U$ the only change to the greens function is the addition of a term that is harmonic 
within the
unit disk.  The parameter of the integration over the position of the vertex operator
would be to the measure on PSL(2,R), giving an infinite factor in the conformally 
invariant
case {\cite{Shatashvili:1993kk,Craps:2001jp,Liu:1988nz}}.
From this argument we have a definite prescription
for the calculation of vertex operator expectation values, which is to use the conformal
transformation to modify the greens function, and calculate the expectation values of 
operators
at the origin with this modified greens function.

\subsection{Emission of one particle in sigma model}

Now we will use this prescription to calculate the sigma model 
expectation values of some
operators, and we will start with the simplest, that of the closed string tachyon.
The vertex operator for the tachyon is $: e^{i p_\mu X^\mu(z(y))
} :$, and it is inserted at the point $y=0$.
The normal ordering prescription for all such operators is
that any divergent
pieces will be subtracted, but finite pieces will remain and by inspection we see that
the appropriate subtraction from the greens function is
\beqn
: {\cal G}^{\mu\nu}( z, z'): &=& G^{\mu\nu}(z,z') - g^{\mu\nu} \alpha' \ln \left| z - z'
\right|
\label{subtractedgf}
\eeqn
Using (\ref{subtractedgf}) we can immediately see that
\beqn
\langle: e^{i p_\mu X^\mu(y=0) } : \rangle &=&
 \left. Z e^{- \frac{1}{2} p_\mu p_\nu : {\cal G}^{\mu\nu}
(z(y), z'(y)) : } \right|_{y=0}
\nonumber \\
&=&
Z \exp \left( - \frac{\alpha'}{2} p_\mu p_\nu
\sum_{n=1}^\infty \left( \frac{g - 2 \pi \alpha' F - \frac{ \alpha'}{2}
\frac{U}{n} }{ g + 2 \pi \alpha' F + \frac{ \alpha'}{2} \frac{U}{n} } \right)^{\mu\nu}
\frac{ 1}{n} \frac{ |b^{2n}|}{|a^{2n}|} \right).
 \label{tachyonEV}
\eeqn
We recall that our proceedure will necessitate an integration over the the parameters of 
the  PSL(2,R) transformation, but comparison with (\ref{1tachyonemissioncalc}) reveals 
that the normalization is fixed by 
\beq
{\cal N} = Z.
\eeq

Having obtained this result fixing the normalization it is natural to push the 
correspondence further as a check of its validity.  We perform a similar analysis 
 for the massless closed string excitations.
In particular the graviton insertion at $y=0$ is given by
\beqn
\langle {\cal V}_h \rangle &=& \langle : - \frac{ 2}{\alpha'}
h_{\mu\nu} \df X^\mu \bar \df X^\nu e^{i p_\mu X^\mu(y=0)} : \rangle
\eeqn
where $h$ is a symmetric traceless tensor and the normalization follows the conventions
of { \cite{Polchinski:1998rq}. } 
This can be analyzed by the same techniques as for the tachyon, noting that there
will be cross contractions between the exponential and the $X$-field prefactors.
Explicitly we obtain
\beqn
\langle {\cal V}_h \rangle &=&
- \frac{ 2}{\alpha'} Z h_{\mu\nu} \left( \df \bar \df' : {\cal 
G}^{\mu\nu}\left(z(y),z'(y)\right) :
+ \df :{\cal G}^{\mu\alpha} \left(z(y), z'(y) \right):
\right.
\nonumber \\
&~& \left. \times \bar \df :{\cal G}^{\mu\beta} \left(z(y), z'(y) \right): (i
p_\alpha) (i p_\beta)
\right)
e^{- \frac{1}{2} p_\mu p_\nu : {\cal G}^{\mu\nu} (z(y), z'(y) ) : }|_{y=0}
\nonumber \\
&=&  
Z h_{\mu\nu}  \left( -
\sum_{n=1}^\infty \left( \frac{g - 2 \pi \alpha' F - \frac{ \alpha'}{2}
\frac{U}{n} }{ g + 2 \pi \alpha' F + \frac{ \alpha'}{2} \frac{U}{n} } \right)^{\mu\nu}
n \frac{|b^{2 (n-1)}| }{|a^{2 (n-1)}| } \frac{1}{|a^2|^2}
\right.
\nonumber \\
&~&  +
\frac{\alpha'}{2}
\sum_{n=1}^\infty \left( \frac{g - 2 \pi \alpha' F - \frac{ \alpha'}{2}
\frac{U}{n} }{ g + 2 \pi \alpha' F + \frac{ \alpha'}{2} \frac{U}{n} } \right)^{\mu\alpha}
\frac{ |b^{2(n-1)} |}{ |a^{2(n-1)} |} \frac{-b}{|a^2|a}
\nonumber \\
&~& \times \left.
\sum_{m=1}^\infty \left( \frac{g - 2 \pi \alpha' F - \frac{ \alpha'}{2}
\frac{U}{m} }{ g + 2 \pi \alpha' F + \frac{ \alpha'}{2} \frac{U}{m} } \right)^{\nu\beta}
\frac{ |b^{2(m-1)} |}{ |a^{2(m-1)} |} \frac{-b*}{|a^2|{a^*}} p_\alpha p_\beta
\right)
\nonumber \\
&~&
\exp \left( - \frac{\alpha'}{2} p_\mu p_\nu
\sum_{n=1}^\infty \left( \frac{g - 2 \pi \alpha' F - \frac{ \alpha'}{2}
\frac{U}{n} }{ g + 2 \pi \alpha' F + \frac{ \alpha'}{2} \frac{U}{n} } \right)^{\mu\nu}
\frac{ 1}{n} \frac{ |b^{2n}|}{|a^{2n}|}  \right).
\nonumber \\
&~&
\label{gravitonEV}
\eeqn
and  by comparing (\ref{gravitonBScalc}) and 
(\ref{gravitonEV}) we see that the relation ${\cal N} = Z$ holds and that the form of the 
two expectation values is indentical in detail.

Finally, we can perform the same kind of calculation for a more general closed string   
state, like the one considered in (\ref{arbstateBScalc}).
We consider a state which may be off shell in the sense that it not annihilated by
the positive modes of the $\sigma$-model energy momentum tensor (the Virasoro
generators), may not satisfy the
mass shell condition, and may
not be level matched.
Our explicit choice is to consider the operator
\beqn
\langle {\cal V}_A \rangle &=&
\langle : -i
\left( \frac{2}{\alpha'} \right)^{3/2}
A_{\mu\nu\delta} \frac{\df^a}{(a-1)!}  X^\mu \frac{\bar
\df^b}{(b-1)! }
X^\nu \frac{\bar \df^c}{(c-1)!} X^\gamma e^{i p_\mu X^\mu } : \rangle
\eeqn
which is an arbitrary state involving three creation operators.
We find that
\beqn
\langle {\cal V}_A \rangle &=&
 Z A_{\mu\nu\delta}
\left( \frac{2}{\alpha'} \right)^{3/2}
\left( \frac{\df^a}{(a-1)!} \frac{\bar \df'^b}{(b-1)!}
: G^{\mu\nu}(z,z') : \frac{\bar \df^c}{(c-1)!} :G^{\delta\alpha}(z,z'): p_\alpha
\right. \nonumber \\
&~& +  \frac{\df^a}{(a-1)!} \frac{\bar \df'^c}{(c-1)!}
: G^{\mu\delta}(z,z') : \frac{\bar \df^b}{(b-1)!} :G^{\nu\alpha}(z,z'): p_\alpha
 \nonumber \\ &~&
\left. -  \frac{\df^a}{(a-1)!} : G^{\mu\alpha}(z,z') :  \frac{\bar \df^b}{(b-1)!}
:G^{\nu\beta}(z,z'): \frac{\bar \df^c}{(c-1)!} :G^{\delta\gamma}(z,z'):
p_\alpha p_\beta p_\gamma \right)
\nonumber \\ &~&
\left. \times e^{- \frac{1}{2} p_\mu p_\nu : {\cal G}^{\mu\nu} (z, z' ) : }
\right|_{y=0}.
\label{arbstateEV}
\eeqn
Comparing (\ref{arbstateEV}) with (\ref{arbstateBScalc}) keeping (\ref{specialcase}) in 
mind it is not hard to see that the two coincide, as they should given the previous 
development.

\subsection{Sigma model emission of many particles}

Having explored the emission of one particle and found many substantial similarities, we now 
look at the emission of two particles.  
We expect that this amplitude will depend upon the relative position of the two vertex operators, and 
since even in the case of conformal invariance there are not enough free parameters to fix two 
closed string vertex operators on the disk world sheet.
We first calculate the expectation value of the emission of two tachyons, with momenta $p$ and $k$.
\beqn
\langle \left. : e^{i k_\mu X^\mu} : \right|_{\omega} 
 \left. : e^{i p_\nu X^\nu } : \right|_0 \rangle
&=& Z \exp \left( - \frac{ k_\mu k_\nu}{2} G^{\mu\nu}\left(z(\omega), z(\omega) \right) \right)
\nonumber \\
&~& \times
  \exp \left( - \frac{ p_\mu p_\nu}{2} G^{\mu\nu} \left( z(0), z(0) \right) \right)
\exp \left( - \frac{ k_\mu  p_\nu}{2} G^{\mu\nu}\left( z(\omega), z(0) \right) \right)
\nonumber \\
&=&
{{\cal A}}_{{\cal T} 2 \sigma }
\eeqn

This is the necessary first step in determining a more arbitrary amplitude.
To make contact with the more complicate amplitudes calculated in (\ref{twovertexbigmess2}) and
(\ref{twovertexbigmess}) we consider
the expression
\beqn
\langle
A_{\mu\nu \delta}  : \frac{ \df^n}{(n-1)!} X^\mu \frac{ \bar \df^p }{(p-1)!} X^\nu
\frac{ \bar \df^q }{(q-1)! } X^\delta e^{i k_\mu X^\mu} : |_\omega 
 : e^{ i p_\mu X^\mu} :|_0 \rangle
\nonumber \\
=
{{\cal A}}_{{\cal T} 2 \sigma }
A_{\mu\nu\delta}
\left[ \left( \frac{ i k_\alpha}{(n-1)!} \df^n G^{\mu\alpha} (z(\omega), z'(\omega) )
+ \frac{i p_\alpha}{(n-1)!} \df^n G^{\mu\alpha} (z(\omega),z'(0) ) \right) \right.
\nonumber \\
\times
\left( \frac{ i k_\beta}{(p-1)!} \bar \df^p G^{\nu\beta}(z(\omega),z'(\omega) )
+ \frac{ i p_\beta}{ (p-1)!} \bar \df^p G^{\nu\beta} ( z(\omega),z'(0) ) \right)
\nonumber \\
\times
\left( \frac{i k_\gamma}{ (q-1)!} \bar \df^q G^{\delta\gamma} (z(\omega),z'(\omega) )
+ \frac{ i p_\gamma}{ (q-1)!} \bar \df^q G^{\delta\gamma}(z(\omega),z'(0) ) \right)
\nonumber \\
+
\frac{ \df^n}{(n-1)!} \frac{\bar \df^{'p}}{(p-1)!}  G^{\mu\nu}(z(\omega),z'(\omega) )
\nonumber \\
\times \left( \frac{i k_\gamma}{ (q-1)!} \bar \df^q G^{\delta\gamma}(z(\omega),z'(\omega) )
+ \frac{ i p_\gamma}{ (q-1)!}\bar \df^q G^{\delta\gamma}(z(\omega),z'(0) ) \right)
\nonumber \\
+
\frac{ \df^n}{(n-1)!}  \frac{\bar \df^{'q}}{(q-1)!} G^{\mu\delta}(z(\omega),z'(\omega) )
\nonumber \\
\times
\left. \left( \frac{ i k_\beta}{(p-1)!} \bar \df^p G^{\nu\beta}(z(\omega),z'(\omega) )
+ \frac{ i p_\beta}{ (p-1)!} \bar \df^p G^{\nu\beta} ( z(\omega),z'(0) ) \right)
\right]
\label{arbstatesigmamodel}
\eeqn
Also note that if we consider 
\beqn
\langle
 : 
e^{i k_\mu X^\mu}  : |_\omega
A_{\mu\nu \delta}  : 
\frac{ \df^n}{(n-1)!} X^\mu \frac{ \bar \df^p }{(p-1)!} X^\nu
\frac{ \bar \df^q }{(q-1)! } e^{ i p_\mu X^\mu} :|_0 \rangle
\nonumber 
\eeqn
we see that it gives the above expression (\ref{arbstatesigmamodel})
with $\omega \leftrightarrow 0$.

To demonstrate the general equivalence of the boundary state approach with that 
of the sigma model the sums that appear in the general expressions of boundary state
matrix elements must be shown to coincide with the expressions that appear above.
To this end consider first the sum that appears in (\ref{twotachyonamp}),
\beqn
\sum_{m=0}^\infty \omega^m M^{(a,b)}_{n m}
&=& \sum_{m=0}^\infty \oint \frac{dz}{2\pi i} \omega^m z^n \frac{ (b^* z + a^*)^{m-1} }{
(a z + b)^{m+1} } 
\nonumber \\
&=&
\oint \frac{dz}{2\pi i}  z^n \frac{1}{ (b^* z + a^*) (a-\omega b^*) } 
\left( z - \frac{a^* \omega - b}{-b^* \omega +a } \right)^{-1}
\nonumber \\
&=&
\left( \frac{ a^* \omega - b}{-b^* \omega + a} \right)^c.
\eeqn
This derivation uses the normalization condition on $a$ and $b$, and can be seen to be equal
to $z^n (y)$ which is the inverse transform of (\ref{conftranfandnorm}).

The other sum that appears generally in this analysis is
\beqn
\sum_{m=0}^{\infty} \frac{m!}{(m-n)!} \omega^{m-n}
M_{rm}^{(ab)}
\nonumber
\eeqn
as seen in (\ref{twovertexbigmess}).  In the case $n>m$ we have used the shorthand 
\beqn 
\frac{m!}{(m-n)!} = m(m-1) \ldots (m-n+1).
\nonumber 
\eeqn
Now consider
\beqn
\sum_{m=0}^{\infty} \frac{m!}{(m-n)!} \omega^{m-n}
M_{rm}^{(ab)}
&=& \sum_{m=0}^{\infty} \oint \frac{dz}{2\pi i} \frac{m!}{(m-n)!} \omega^{m-n}
z^r \frac{ (b^* z + a^*)^{m-1} }{ (az + b)^{m+1} }
\nonumber \\
&=& n! \oint \frac{dz}{2\pi i} z^r \frac{ (b^* z + a^* )^{n-1} }{ (a-b^* \omega)^{n+1} }
\left( z - \frac{ a^* \omega - b }{- b^* \omega + a} \right)^{-n-1}
\nonumber \\
&=& \left. \df^n z^r (b^* z + a^*)^{n-1} (a-b^* \omega)^{-(n+1)} \right|_{z = 
\frac{ a^* \omega - b }{- b^* \omega + a} }
\nonumber \\
&=& \left. \df^n \left( \frac{ a^* z - b}{ -b^* z + a } \right)^r \right|_{z=\omega}.
\label{inductionneeded}
\eeqn
The last equality in this can be shown by induction, the case for $n=1$ is trivial, and 
so we demonstrate the induction.
First note that use of the Leibnitz rule gives 
\beqn
\left. \df^k z^n (b^* z + a^*)^{k-1} (a-b^* \omega)^{-(k+1)} \right|_{z =
\frac{ a^* \omega - b }{- b^* \omega + a} }
=
\nonumber \\
\frac{ (a^* \omega - b)^{n-k} }{ (- b^* \omega + a )^{n+k} } 
 \sum_{j=0}^k \left( \matrix{ k \cr j } \right) n \ldots (n -(k-j-1) ) (k-1) 
\ldots (k-j) b^{*j} (a^* \omega - b)^j.
\eeqn
Now consider for $k=k_0 +1$ the expression can be manipulated
\beqn
\df^k 
\left. \left( \frac{ a^* z - b}{ -b^* z + a } \right)^n \right|_{z=\omega}
&=& \df \left( \df^{k_0} \left( \frac{ a^* z - b}{ -b^* z + a } \right)^n \right)
\nonumber \\
&=& \df \frac{ (a^* z- b)^{n-k_0} }{ (- b^* z+ a )^{n+k_0} }
 \sum_{j=0}^{k_0} \left( \matrix{ k_0 \cr j } \right) n \ldots (n -(k_0-j-1) ) 
\nonumber \\ &~& ~~~~ (k_0-1)
\ldots (k_0-j) b^{*j} (a^* z - b)^j
\nonumber \\
&=& \frac{ (a^* z- b)^{n-(k_0+1)} }{ (- b^* z+ a )^{n+(k_0+1) } }
\sum_{j=0}^{k_0} \left( \matrix{ k_0 \cr j } \right) n \ldots (n -(k_0-j-1) )
\nonumber \\ &~& ~~~~ (k_0-1)
\ldots (k_0-j) b^{*j} (a^* z - b)^j
\left( (n+j-k_0) + (2 k_0 -j) b^* (a^*z - b) \right)
\nonumber \\
&=&
\frac{ (a^* z- b)^{n-(k_0+1)} }{ (- b^* z+ a )^{n+(k_0+1) } }
\sum_{j=0}^{k_0+1} \left( \matrix{ k_0+1\cr j } \right) n \ldots (n -(k_0-j) )
\nonumber \\ &~& ~~~~
k_0
\ldots (k_0+1-j) b^{*j} (a^* z - b)^j
\eeqn
as desired.
This demonstrates the induction step, and the validity of ({\ref{inductionneeded}}).
Note that similar results can be obtained for expressions with negative indices on $M_{nk}^{(a,b)}$
and negative powers of $\omega$.  These are obtained from considering the boundary state on the 
right of the matrix elements.  This have to be interpreted as a dual description of the boundary
states presented.  This is because radial quantization and the operator state correspondence imply
that in this case the domain of interest is the complex plane with the unit disk excluded.  This is
equally a fundamental region of the plane, and the conformal transformation between the two is
$\omega \rightarrow \frac{1}{\bar \omega}$, a fact which is intimated at by the fact that
$M_{-n-k}^{(a,b)} = M_{nk}^{*(a,b)}$.

Now we have demonstrated that the results obtained from the boundary state calculations
exactly match those of the sigma model after the propagator including the boundary perturbations
has been obtained, and the resulting expression has been transformed into a new coordinate system.
This shows that the boundary state algebraizes all matrix elements that would otherwise be
calculated in the sigma model.  This observation will be important as we generalize these results
to higher genus surfaces.  We also remark that the result explicitly presented for 
the emission of two closed string states clearly generalizes to the emission of an arbitrary
number of such particles.  Mechanically this can be seen because the commutation of two such
vertex operators to produce a normal-ordered expression produces the familiar logarithmic
term, and the boundary state gives the $F$ and $U$ dependence within the inner product.

\section{Higher Genus amplitudes}

Now that the overlap of the boundary state with either single or multiple particle
states we have the tools that are needed to proceed and determine higher order contributions 
in the string loop sense to the vacuum energy of the object described by the boundary
state.  We will proceed in the following manner, by utilizing a sewing construction to
relate higher order amplitudes to products of lower order amplitudes.  
The procedure outlined as envisioned can produce an arbitrary number of interactions 
with the boundary state at the oriented tree level, and an arbitrary number of handles
and interactions with the boundary state in the unoriented sector.  As is well known the 
description of higher genus orientable surfaces is a more difficult subject and the 
constuction will produce results that are implicit rather than explicit.
The final result will be several terms in the Euler number expansion so that
\beqn
{\cal Z} = Z_{disk} + Z_{annulus} + Z_{Mobius Strip} + \ldots
\eeqn
with each term carrying the appropriate power of the open string coupling constant.  
The results in this section will be organized by Euler number, and where appropriate
compared with other similar results in the literature.

\subsection{$\chi=1$}

There are two surfaces with $\chi=1$, the disk and $RP^2$.  The non-orientable surface 
$RP^2$, see {\cite{Itoyama:2002dd}} for details in another context, has no interaction with 
the fields $F$ and $U$ and so is not of interest for this analysis.  The disk by contrast 
has been analysed previously in this work and the contribution to the partition function
for the boundary state is given by its overlap with the unit operator (equivalently the 
tachyon with zero momentum), as given in (\ref{diskpf}).

\subsection{$\chi=2$}

There are several surfaces that have an Euler number of 2.  The easiest to discuss in this
is the torus, which is immaterial for the same reason that $RP^2$ was among the surfaces
with $\chi=1$, namely that it has no interactions with $F$ or $U$.
Similarly the Klein bottle, the unoriented equivalent of the torus, will not contribute to
the partition function.
We are left with the annulus and with the Mobius strip, as the nontrivial contriubutions at
this level.  The annulus can be thought of as the tree level closed string exchange
chanel.  The Mobius strip is the non-orientable analogue of the disk.

Considering first the annulus that was considered in detail in {\cite{Akhmedov:2001yh}}, we 
recapitulate some of the salient results.
Suppressing for notational simplicity the integrations over the parameters of the conformal
transformations we have that
\beqn
 Z_{annulus} = \langle B_{a,b} | \frac{1}{\Delta} | B_{a',b'} \rangle.
\eeqn
Using the integral representation of the closed string propagator
\beqn
\frac{1}{\Delta} = \frac{1}{4\pi} \int \frac{d^2 z}{|z|^2} z^{L_0 -1} \bar z^{L_0 -1}
\nonumber 
\eeqn
and suppressing the integrals and, temporaraly, the zero modes for notational clairity we have
\beqn
 Z_{annulus}
&=& Z_{disk}^2 \langle 0 | \exp{  \left(-\alpha^\mu_i M^{(a,b)}_{ni} \Lambda^n_{\mu\nu}
M^{(a,b)*}_{nj} \tilde \alpha^\nu_j \right) } 
z^{\sum \alpha_{-n} \alpha_n} \bar z^{\sum \tilde \alpha_{-n} \tilde \alpha_n}
\nonumber \\
&~&
\exp{ \left( - \alpha^\gamma_{-k} M^{(a',b')}_{-m-k} \Lambda^m_{\gamma\delta}
M^{(a',b')*}_{-m-l} \tilde \alpha^\delta_{-l} \right)  } | 0 \rangle
\nonumber \\
&=& Z_{disk}^2 \langle 0 | \exp{ \left( -\alpha^\mu_i M^{(a,b)}_{ni} \Lambda^n_{\mu\nu}
M^{(a,b)*}_{nj} \tilde \alpha^\nu_j \right) }
\nonumber \\
&~&
\exp{\left( -z^k \alpha^\gamma_{-k} M^{(a',b')}_{-m-k} \Lambda^m_{\gamma\delta}
M^{(a',b')*}_{-m-l} \tilde \alpha^\delta_{-l} \bar z^l \right)  } | 0 \rangle
\nonumber \\
&=&  \exp \left( \sum_{k=1}^\infty \frac{1}{k} g^{\mu\nu} \delta_{rs}
\left\{ \left[ r M^{(a,b)}_{nr} \Lambda^n_{\mu\alpha} M^{(a,b)*}_{nj} j \bar z^j
g^{\alpha \delta} M^{(a',b')*}_{-m-j} \Lambda^m_{\nu\delta} M^{(a',b')}_{-m-s} z^s
\right]^k \right\}^{rs}_{\mu\nu} \right)
\nonumber \\
&~& \times Z_{disk}^2 F(p).
\label{annulusbyBS}
\eeqn
Verifying the first equality requires that one use the Baker-Hausdorf formula for 
commutators of exponentials, and the second equality is an application of Wick's theorem.
The term in the last exponential is understood to have its powers defined by contraction
of both the lorentz and oscillator indices, and the number $F(p)$ is a gaussian factor 
dependant on the (otherwise implicit) momentum of each boundary state, which can be read 
off from the boundary conditions (\ref{conditiononP}).
Explicitly the form of $F(p)$ is given by
\beqn
 F(p) &=& \exp \left\{ p^\mu p^\nu \left[ \left( 
\delta_{0j} g_{\mu\delta}- M^{(a,b)}_{n0} 
\Lambda^n_{\mu\delta}
M^{(a,b)*}_{nj} \right) \right. \right.
\nonumber \\
&~&
\left( \frac{1}{ \delta_{jk} g_{\delta\gamma} - j \bar z^j M^{(a',b')*}_{-m-j} \Lambda^m_{\eta\delta}
M^{(a',b')}_{-m-l} g^{\eta\zeta} z^l l M^{(a,b)}_{rl} \Lambda^r_{\zeta\gamma} M^{(a,b)*}_{rk} }
\right)^{\delta\gamma}_{jk} 
\nonumber \\
&~&
\left. \left. \left( \delta_{k0} g_{\gamma\nu} - k \bar z^k M^{(a',b')*}_{sk} \Lambda^s_{\nu \gamma}
M^{(a',b')}_{s0} \right) - g_{\mu\nu} \right] \right\}.
\eeqn
In addition this is multiplied by terms coming from the zero mode part of the propagator.  
In the preceeding 
equations the oscillator index has been chosen as positive or zero to
make the negative signs meaningful.  In all cases, repeated indices indicate summation.

We note in passing that the cases of $U \rightarrow 0$ and $U \rightarrow \infty$ give a particularly
simple form for the matrices $M \Lambda M^*$.  
We have
\beqn
\left. M^{(a,b)}_{k m} \Lambda^k_{\mu\nu} M^{(a,b)*}_{kn} 
\right|_{U \rightarrow 0} &=& M^{(a,b)}_{k m} \frac{1}{k} \left( \frac{ g - 2\pi \alpha' F}
{g+ 2\pi \alpha' F} \right)_{\mu\nu} M^{(a,b)*}_{kn}
\nonumber \\
&=& \left( \frac{ g - 2\pi \alpha' F}
{g+ 2\pi \alpha' F} \right)_{\mu\nu} \frac{1}{m} \delta_{mn},
\eeqn
and similarly
\beqn
\left. M^{(a,b)}_{k m} \Lambda^k_{\mu\nu} M^{(a,b)*}_{kn}
\right|_{U \rightarrow \infty}  &=&  -g_{\mu\nu} \frac{1}{m} \delta_{mn}.
\eeqn
These results can be obtained by explicit contour integration using the definition of
$M$.  We can see that the $U=0$ case gives the boundary state of a background gauge field
{\cite{Lee:2001ey}} and when $U=\infty$ a localized object appears.  In fact this parameter
$U$ interpolates between Neumann and Dirichlet boundary conditions {\cite{Kraus:2000nj}}.

It is worthwhile to check the result obtained in (\ref{annulusbyBS}) in the known case where only 
the field $F$ is present.  Then the boundary conditions enforce that $p=0$, and with the above
simplification we find
\beqn
Z_{annulus}(F) &=& 
\exp \left( \sum_{k=1}^\infty \frac{1}{k} g^{\mu\nu} \delta_{rs}
\left\{ \left[ r \delta_{rj} \left( \frac{g - 2\pi\alpha'F}{g + 2\pi\alpha' F}\right)_{\mu
\alpha} \frac{1}{j} g^{\alpha\delta}
\right. \right. \right.
\nonumber \\
&~& \left. \left. \left. ~~~~~ j \bar z^j \delta_{js} \left( \frac{g - 2\pi\alpha'F}{g + 2\pi\alpha' 
F}\right)_{\delta\nu}
 z^s
\right]^k \right\}^{rs}_{\mu\nu} \right)
 Z_{disk}^2 F(0)
\nonumber \\
&=& \exp \left( - \sum_{r=0}^\infty Tr \ln \left( g - \frac{1 + |z|^{2r} }{1- |z|^{2r} }
4 \pi \alpha' F + 4 \pi^2 \alpha'^2 F^2 \right) 
\right. 
\nonumber \\
&~&
\left. - \sum_{r=0}^\infty Tr \ln \left( g (1 - |z|^{2r})
\right) - \frac{1}{2} Tr \ln \left( g + 4 \pi \alpha' F + 4 \pi^2 \alpha'^2 F^2 \right) \right)
Z_{disk}^2
\nonumber \\
&=&
\prod_{r=1}^\infty (1 - |z|^{2r})^{-D} \prod_{r=1}^\infty \det \left( 
g - \frac{1 + |z|^{2r} }{1- |z|^{2r} }
4 \pi \alpha' F + 4 \pi^2 \alpha'^2 F^2 \right)^{-1}. 
\eeqn
This result agrees upon the inclusion of the ghost contribution with that obtained 
in {\cite{Fradkin:1985qd}}.  Note that the partition function for the disk 
is cancelled by the term constant in $r$ which is then summed using $\zeta$ function
regularization, exactly as the Born-Infeld action was obtained in the first place.

In a similar method we can obtain the partition function for the Mobius strip in this 
background as well.  We use the cross cap state elaborated on in {\cite{Itoyama:2002dd} }.
\beqn 
| {\cal C} \rangle = \exp \left( - \sum_{n=1}^\infty \frac{ (-1)^n}{n} \alpha_{-n} \tilde
\alpha_{-n} \right) | 0 \rangle.
\eeqn
Using this in analogy with the development of (\ref{annulusbyBS}) we find that the 
\beqn
 Z_{mobius}
&=& \langle B_{a,b} | \frac{1}{\Delta} | {\cal C} \rangle
\nonumber \\
&=& Z_{disk} Z_{ RP^2 }
 \exp \left( \sum_{k=1}^\infty \frac{1}{k} g^{\mu\nu} \delta_{rs}
\left\{ \left[ r M^{(a,b)}_{nr} \Lambda^n_{\mu\nu} M^{(a,b)*}_{nj} j \bar z^j
 \frac{ (-1)^s }{s} \delta_{js} z^s
\right]^k \right\}^{rs}_{\mu\nu} \right).
\label{mobiusbyBS}
\nonumber \\
\eeqn
As in the case of the annulus, we find that the contributions $Z_{disk}$
cancel explicitly when we go to the $U=0$ limit, 
where conformal invariance
is restored.  In that limit we find
\beqn
 Z_{mobius}(F) 
&=& 
\prod_{m=1}^{\infty} \det \left( g + F \frac{1 + (-1)^m |z|^{2m} }{ 1 - (-1)^m |z|^{2m} }
\right)^{-1}.
\eeqn

Finally it is amusing to check and make sure that an analogous calculation will go 
through and reproduce the known partition function for the Klein bottle.
Instead of the two copies of the boundary state two cross caps are inserted, and the resulting 
expression 
\beqn
Z_{K^2} &=& Z_{RP^2}^2 \exp \left( -\sum_{r=1}^\infty g^{\mu\nu} \ln \left( g_{\mu\nu}
(1 - |z|^{2r}) \right) \right)
\eeqn
which can be seen to reduce to Dedekind $\eta$-functions, in agreement with the known result
{\cite{Polchinski:1998rq}}.

\subsection{ $\chi = 3$}
To extend it beyond $\chi=2$ the boundary state formalism requires careful contemplation.
We propose the following method which allows the construction of states of arbitrary 
Euler number, and for the nonorientable sector in principle a complete description of the
dynamics.  The procedure proposed is as follows; using the sewing construction for higher genus
amplitudes (described in \cite{Polchinski:1998rq} among others)
 and armed with the result proved 
earlier in this paper that the emission 
of any particle from the bosonic boundary state corresponds with the expectation of a
vertex operator inserted at a definite position on the disk, we propose to add any number of
interactions with the brane described by the boundary state and any number of cross caps.

To recapitulate, the idea of the sewing construction is to create a higher genus amplitude
by joining two lower genus amplitudes by inserting a vertex operator on each of the lower genus amplitudes
and summing over the vertex operator.
Explicitly the construction is
\beqn
\langle : A_1 : \ldots :A_n: \rangle_{M} &=& \int_\omega \sum_{V} \omega^{h_V} \bar \omega^{\tilde h_V}
\langle :A_1: 
\ldots :V: \rangle_{M_1}
\langle :V: \ldots :A_n: \rangle_{M_2}
\eeqn 
with $M= M_1 \# M_2$ and $\ldots$ represents arbitrary vertex insertions.
This construction is tantamount to adding a closed string propagator between the two manifolds with 
vertex operators on them.
Since we have shown that the emission of one particle from the disk with $F$ and $U$ on its
boundary matches the overlap obtained from the boundary state
\beqn
\langle V | B_{a,b} \rangle &=& \langle :V: \rangle_{T_0,U,F}
\eeqn
we can then use this to obtain the contribution of a boundary with the fields
$U$ and $F$ at it.
This sort of construction was considered in {\cite{Callan:1987px}}.

The novel feature presented here is the generalization of the boundary state and cross-cap operators
through the state operator correspondence.  The fact that sphere amplitude for three string scattering
is conformally invariant is used, in combination with the fact that both $|{\cal C} \rangle$ and 
$|B_{a,b} \rangle$ both have a well defined overlap with any closed string state allows us to take
the expression 
\beqn
\frac{1}{\Delta} | B_{a,b} \rangle &=& \int \frac{dz d\bar z}{|z|^4} |z|^{p^2}
\exp{\left( -z^k \alpha^\gamma_{-k} M^{(a',b')}_{-m-k} \Lambda^m_{\gamma\delta}
M^{(a',b')*}_{-m-l} \tilde \alpha^\delta_{-l} \bar z^l \right)  } | 0 \rangle
\eeqn
and its equivalent using $|{\cal C} \rangle$ to 
(suppressing prefactors)
\beqn
\exp{\left( -z^k \frac{\df^k}{(k-1)!} X^\gamma  M^{(a',b')}_{-m-k} \Lambda^m_{\gamma\delta}
M^{(a',b')*}_{-m-l} \frac{\bar \df^l}{(l-1)!} X^\delta \bar z^l \right)  } 
\label{Boperatorosc}
\eeqn
by use of the operator state correspondence.  These states are inserted within 
expectation values to give higher genus contributions.

There are several different states with $\chi = 3$.  The most obvious are the four possible
insertions of boundary states and cross caps, and the addition of a handle to either a boundary
state or cross cap (thereby increasing from $\chi =1$ to $\chi =3$ because increasing the genus
by $1$ increases the Euler number by $2$.  Note that the state with three cross caps and the 
state with a cross cap and a handle are topologically equivalent.

To obtain the amplitude for three boundaries we calculate
\beqn
Z_{'pants'} &=& \langle B_{a,b} | \frac{1}{\Delta} :B_{a',b'}: \frac{1}{\Delta} |B_{a'',b''} \rangle
\nonumber 
\eeqn
where $:B_{a',b'}:$ is as given in (\ref{Boperatorosc}
). 
Noting that the coefficient of $\alpha_m$ in $\frac{\df^n}{(n-1)! } X$ is
\beqn
\frac{\df^n}{(n-1)! } X &=& \sum_{a=-\infty}^\infty D_{na} \alpha_a,
 \\
D_{na} &=& (-1)^{n-1} \frac{ (a+1) \ldots (n+a-1) }{(n-1)! },
\eeqn
we proceed to calculate
\beqn
Z_{'pants'} &=& Z_{disk}^3 
F_0 \left(p \right)
 \exp \left( \sum_k \frac{1}{k} \delta_{na}
\left( n C_{nm}(1) m C_{am}(3)
\right)^k \right)
\nonumber \\
&~&
\exp \left( \sum_k \frac{1}{k} \delta_{na} 
\left( n C_{nm}(1) m D_{n' -a} C_{n'm'}(2) 
\bar D_{m' -m} \right)^k \right)
\nonumber \\
&~& \exp \left( \sum_k \frac{1}{k} \delta_{na} 
\left( n C_{nm}(3) m D_{n' a} C_{n'm'}(2)
\bar D_{m' m} \right)^k \right) 
\nonumber \\
&~& \exp \left( \sum_k \frac{1}{k} \delta_{na}
\left(  n C_{nm}(1) m D_{n' j} C_{n'm'}(2)
\bar D_{m' -m} j C_{jk}(3) k D_{n'' -a} C_{n''m''}(2)
\bar D_{m'' k} \right)^k \right)
\nonumber \\
\eeqn
Where as in (\ref{annulusbyBS}) $F_0 \left(p \right)$ is a complicated function which is
gaussian in the momentum of the boundary state, the integrals are implicit, and the expression
$C_{nm}(i)$ is an abbreviation
\beqn
C_{nm}(i) &=& z_i^n M^{(a,b)}_{kn} \Lambda^k_{\mu\nu} M^{(a,b)*}_{km} \bar z_i^m
\eeqn
with $i$ an index reminding which integration from the closed string propagator $z_i$ came from.

From this we see immediately that 
the contributions for the genus expansion become increasingly 
complicated as 
$\chi$ increases.  In the particularly simple case of a vanishing 
tachyon, this can be evaluated and one
obtains a product of exponentials of hypergeometric functions.  
In particular for the case of the constant $F$ field we obtain
\beqn
Z_{'pants'}(F) &=& Z_{disk}^3
\exp \left( - \sum_n Tr \ln \left( 1 - |z_1 z_3|^{2n} \left( \frac{g-2\pi\alpha' F}{g+2\pi\alpha' F}
\right)^2 \right) \right)
\nonumber \\
&~& 
\exp \left( - \sum_{na} Tr \ln \left( 1 - n |z_1|^{2n} |z_2|^2 F(-n+1,-a+1;2;|z_2|^2) 
 \left( \frac{g-2\pi\alpha' F}{g+2\pi\alpha' F}
\right)^2 \right) \right)
\nonumber \\
&~&
\exp \left( - \sum_{na} Tr \ln \left( 1 - n |z_3|^{2n} |z_2|^2 F(n+1,a+1;2;|z_2|^2)
 \left( \frac{g-2\pi\alpha' F}{g+2\pi\alpha' F}
\right)^2 \right) \right)
\nonumber \\
&~&
\exp \left( - \sum_{nma} Tr \ln \left( 1 - n |z_1|^{2n} |z_2|^2 F(-n+1,m+1;2;|z_2|^2)
\right. \right. \nonumber \\
&~& \left. \left. m |z_3|^{2m} |z_2|^2 F(m+1,-a+1;2;|z_2|^2)
 \left( \frac{g-2\pi\alpha' F}{g+2\pi\alpha' F}
\right)^2 \right) \right).
\eeqn
In the above $F(a,b;c;x)$ is the hypergeometric function defined by its series expansion
\beqn
F(a,b;c;x) &=& \sum_{n=0}^\infty \frac{ (a+n-1)! (b+n-1)! (c-1)! }{ n! (a-1)! (b-1)! (c+n-1)! }
x^n,
\eeqn
and the logarithm is interpreted as its series expansion, and both lorentz and oscillator 
indices are summed over.
Note that this expression has many of the properties that we expect for the partition function
on a twice punctured disk.  In particular this depends on three parameters (the $z_i$ terms arising
from the integration over the propagators to the various boundary states) which can be 
identified as the Teichmuller parameters for this surface.  
In the limit of any of these parameters going to zero the dominant contribution is from the 
annulus amplitude.  The 
analogous amplitude with any number of cross-caps gives a similar 
expression with the following modifications, for each cross-cap the arguement in the 
hypergeometric expression acquires a negative sign, and the corresponding matrix of 
lorentz indices undergoes the substitution $\frac{ g- 2\pi \alpha' F}{g+2\pi\alpha' F}
\rightarrow g$.

The other two diagrams that must be calculated are the corrections to the disk and to $RP^2$ which
come from the addition of a handle.  This addition is achieved by taking the trace, weighted by 
a factor exponentiated to the level number (coming from the propagator within the handle), 
which is an 
identical operation to taking the expectation 
value of this operator on the torus.
For this calculation it is necessary to take the trace of an operator that 
generically has the normal ordered form
\beqn
:\exp \left( - \alpha_n {\cal M}_{nm} \tilde \alpha_m \right) :
\nonumber 
\eeqn
where the indices on ${\cal M}$ can be either positive or negative, with ${\cal M}$ defined
by
\beqn
{\cal M}_{mn} &=& D_{n'm} C_{n'm'} \bar D_{m'n}.
\eeqn
After a considerable amount of algebra we find by summing over all states in 
the Fock space that
\beqn
Tr \left(
\omega^h \tilde \omega^{\tilde h} :\exp \left( - \alpha_n {\cal M}_{nm} \tilde \alpha_m 
\right) : 
\right)
&=& \prod_{n=1}^\infty \frac{1}{\left| 1-\omega^{h_i} \right|^2 } 
\prod_{n=1}^\infty \frac{ 1}{ 1 - \left( 
\frac{ |a| \omega^{h_a} }{ 1- \omega^{h_a} }
{\cal M}_{ab} \frac{ |b| \tilde \omega^{\tilde h_b} }{ 1 - \tilde \omega^{\tilde h_b} }  
{\cal M}_{-c-b}  \right)^n }. 
\nonumber \\
\label{trace}
\eeqn
This expression uses the convention that the sums within the denominator run over positive
and negative indices.  This suppresses the contribution from the momentum of the loop which is
given by a gaussian, explicitly
\beqn
F(p) &=&  \exp \left\{ p p \left[ \left(
\delta_{0j} - {\cal M}_{0j} \frac{ |j| \tilde \omega^{h_j} }{ 1- \tilde \omega^{h_j} } 
\right) 
\left( \frac{1}{ \delta_{kj} - 
{\cal M}_{kj} \frac{ |k| \omega^{h_k} }{ 1- \omega^{h_k} }
{\cal M}_{kl} } \right)
 \left( \delta_{0l}  - {\cal M}_{0l} 
\frac{ |l| \tilde \omega^{h_l} }{ 1- \tilde \omega^{h_l} } \right) -1  \right] \right\}.
\nonumber \\
\eeqn
The specialization to the case of only interactions with a background $F$ field is given by
substituting $|z|^2 F(a+1,b+1;2;|z|^2)$ for ${\cal M}_{ab}$.

It is interesting at this point to compare the results for this procedure with those
obtained by the standard method of constructing the Greens function on an arbitrary surface
\cite{Schiffer:1954}, and then integrating out the boundary interaction as described previously
(\ref{bosonicgf}).  The Greens function of a unit disk with Neumann boundary conditions with 
a puncture of radius $\epsilon$ at $z=0$ and a puncture of radius $\delta$ at $z = r e^{i \psi}$
is given by 
\beqn
G'(z,z') &=& G(z,z') + \left( \ln \epsilon \right)^{-1} G(z, 0) G(z',0)
+  \left( \ln \delta\right)^{-1} G(z, r e^{i \psi}) G(z', r e^{i \psi})
\nonumber \\
&~&
- Re  \left( 4 \delta^2 \left( \frac{1}{ z - r e^{i \psi} } + \frac{\bar z}{1- \bar z r e^{i \psi} }
\right) \left(  \frac{1}{ \bar z' - r e^{-i \psi} } + \frac{z'}{1-z' r e^{-i \psi} } \right) \right)
\nonumber \\
&~& - Re \left( 4 \epsilon^2 \left( z^{-1} + \bar z \right) \left( \bar z'^{-1} + z' \right)
\right) + O(\epsilon^2) + O(\epsilon^2 \delta^2) + O(\delta^2).
\eeqn
In the above the explicit form of the greens function for the disk (\ref{diskprop}) has been 
substituted into the last two lines.
Integrating out the background field $F$ can be done by recasting this as a one dimensional 
$3\times3$ matrix model.  When this is done the interaction with a field on the boundary
can be integrated out, much as was done for the $2 \times 2$ case in \cite{Fradkin:1985qd}, and
the resulting expression contains the lowest order terms (in the teichmuller parameter) 
of the hypergeometric  functions obtained previously.  Similarly there is  a procedure for obtaining
the Greens function for the disk with a handle added between balls of radius 
$\epsilon$ centered at $z=0$ and $z=r e^{i \psi}$.  This
gives
\beqn
G'(z,z') &=& G(z,z') + \left( \ln \epsilon \right)^{-1} \left( G(z,0) - G(z,r e^{i \psi}) \right)
\left( G(z',0) - G(z',r e^{i \psi}) \right)
\nonumber \\
&~&  - Re \left[ 4 \epsilon^2 \left( z^{-1} + \bar z \right) 
\left( \frac{1}{ z' - r e^{i \psi} } + \frac{\bar z'}{1- \bar z' r e^{i \psi} }
\right)
\right.
\nonumber \\
&~& \left.
+  \left( z'^{-1} + \bar z' \right) 
\left( \frac{1}{ z - r e^{i \psi} } + \frac{\bar z}{1- \bar z r e^{i \psi} }
\right) \right] + O(\epsilon^2) .
\eeqn
As in the case of the disk with holes removed, this greens function can be then used to integrate 
out the quadratic purturbation, obtaining results that are consistant with those presented 
in (\ref{trace}).

\subsection{ $\chi = 4$}

As for $\chi =3$ there are a number of different surfaces of this genus that can be obtained with
the insertion of handles, cross-caps, and boundaries.  The method presented above provides a concrete
proposal for the construction of these higher genus amplitudes for all $\chi \ge 3$.  The construction
is particularly appropriate for what can be interpreted as tree level scattering amplitudes for 
an arbitrary number of closed strings emitted from the brane described by the boundary state.

\section{Discussion and Conclusion}

In this paper we have further explored the boundary state formalism \cite{Callan:1987px}
 and discussed its 
extension to the off-shell case including interaction with a tachyon field of quadratic
profile.  The boundary state has been shown to reproduce the $\sigma$ model calculations 
for emission of any number of closed string states, as detailed in the correspondence
\beqn
\langle V_1 | :V_2: \ldots |B_{a,b} \rangle &=& \langle :V_1: :V_2: \ldots \rangle_{T_0,U,F}.
\eeqn
This can be restated as the fact that the boundary state algebraizes the bosonic string 
propagator.
It has been shown that the inner product of two of the boundary states also reproduces the
$\sigma$ model calculations for a worldsheet of the appropriate genus.
We also present a generalization of this to higher genus, the results of which become progressively
more complicated.
In the case of vanishing tachyon field we obtain the following  expansion in the 
open string coupling constant $g_o$
\beqn
Z_{F} &=& \sum_\chi g_o^\chi Z_\chi \nonumber \\
&=& g_o^{-1} \sqrt{ \det \left( g + 2 \pi \alpha' F \right) }
\nonumber \\
&~& + \int \prod_r \left( 1 - |z^2|^r \right)^{-D} \prod_r \det \left(
g- \frac{1+ |z^2|^r}{1-|z^2|^r} 4 \pi\alpha' F + 4 \pi^2 \alpha'^2 F^2 \right)^{-1}
\nonumber \\
&~& + \int \prod_r \left( 1 - (-1)^r |z^2|^r \right)^{-D} \prod_r \det \left(
g- \frac{1+ (-1)^r |z^2|^r}{1-(-1)^r|z^2|^r} 2\pi \alpha' F \right)^{-1}
\nonumber \\
&~& +
g_o \int \exp \left( - \sum_n Tr \ln \left( 1 - |z_1 z_3|^{2n} \left( \frac{g-2\pi\alpha' 
F}{g+2\pi\alpha' F}
\right)^2 \right) \right)
\nonumber \\
&~&
\exp \left( - \sum_{na} Tr \ln \left( 1 - n |z_1|^{2n} |z_2|^2 F(-n+1,-a+1;2;|z_2|^2)
 \left( \frac{g-2\pi\alpha' F}{g+2\pi\alpha' F}
\right)^2 \right) \right)
\nonumber \\
&~&
\exp \left( - \sum_{na} Tr \ln \left( 1 - n |z_3|^{2n} |z_2|^2 F(n+1,a+1;2;|z_2|^2)
 \left( \frac{g-2\pi\alpha' F}{g+2\pi\alpha' F}
\right)^2 \right) \right)
\nonumber \\
&~&
\exp \left( - \sum_{nma} Tr \ln \left( 1 - n |z_1|^{2n} |z_2|^2 F(-n+1,m+1;2;|z_2|^2)
\right. \right. \nonumber \\
&~& \left. \left. m |z_3|^{2m} |z_2|^2 F(m+1,-a+1;2;|z_2|^2)
 \left( \frac{g-2\pi\alpha' F}{g+2\pi\alpha' F}
\right)^2 \right) \right) 
\nonumber \\
&~& +
g_o
 \int \exp \left( - \sum_n Tr \ln \left( 1 - |z_1 z_3|^{2n} \left( \frac{g-2\pi\alpha' 
F}{g+2\pi\alpha' 
F}
\right)^2 \right) \right)
\nonumber \\
&~&
\exp \left( - \sum_{na} Tr \ln \left( 1 + n |z_1|^{2n} |z_2|^2 F(-n+1,-a+1;2;-|z_2|^2)
 \left( \frac{g-2\pi\alpha' F}{g+2\pi\alpha' F}
\right) \right) \right)
\nonumber \\
&~&
\exp \left( - \sum_{na} Tr \ln \left( 1 - n |z_3|^{2n} |z_2|^2 F(n+1,a+1;2;-|z_2|^2)
 \left( \frac{g-2\pi\alpha' F}{g+2\pi\alpha' F}
\right) \right) \right)
\nonumber \\
&~&
\exp \left( - \sum_{nma} Tr \ln \left( 1 - n |z_1|^{2n} |z_2|^2 F(-n+1,m+1;2;-|z_2|^2)
\right. \right. \nonumber \\
&~& \left. \left. m |z_3|^{2m} |z_2|^2 F(m+1,-a+1;2;-|z_2|^2)
 \left( \frac{g-2\pi\alpha' F}{g+2\pi\alpha' F}
\right)^2 \right) \right)
\nonumber \\
&~& + 
g_o \int \exp \left( - \sum_n Tr \ln \left( 1 - (-1)^n 
|z_1 z_3|^{2n} \left( \frac{g-2\pi\alpha' 
F}{g+2\pi\alpha'
F}
\right) \right) \right)
\nonumber \\
&~&
\exp \left( - \sum_{na} Tr \ln \left( 1 + n |z_1|^{2n} |z_2|^2 F(-n+1,-a+1;2;-|z_2|^2)
 \left( \frac{g-2\pi\alpha' F}{g+2\pi\alpha' F}
\right) \right) \right)
\nonumber \\
&~&
\exp \left( - \sum_{na} Tr \ln \left( 1 - n (-1)^n |z_3|^{2n} |z_2|^2 F(n+1,a+1;2;-|z_2|^2)
 \right) \right)
\nonumber \\
&~&
\exp \left( - \sum_{nma} Tr \ln \left( 1 - n |z_1|^{2n} |z_2|^2 F(-n+1,m+1;2;-|z_2|^2)
\right. \right. \nonumber \\
&~& \left. \left. m (-1)^m |z_3|^{2m} |z_2|^2 F(m+1,-a+1;2;-|z_2|^2)
 \left( \frac{g-2\pi\alpha' F}{g+2\pi\alpha' F}
\right) \right) \right)
\nonumber \\
&~& +
g_o \int \prod_{n=1}^\infty \frac{1}{\left| 1-\omega^{h_i} \right|^2 }
\nonumber \\
&~& \prod_{n=1}^\infty \frac{ 1}{ 1 - \left(
\frac{ |a| \omega^{h_a} }{ 1- \omega^{h_a} }
|z|^2 F(a+1,b+1;2;|z|^2)
 \frac{ |b| \tilde \omega^{\tilde h_b} }{ 1 - \tilde \omega^{\tilde h_b} }
|z|^2 F(-c+1,-b+1;2;|z|^2) \left( \frac{g - 2\pi \alpha' F}{g+2 \pi \alpha'F} \right)^2
  \right)^n 
}
\nonumber \\
&~& 
\exp \left\{ p p \left[ \left(
\delta_{0j} - |z|^2 F(1,j+1;2;|z|^2)  \left( \frac{g - 2\pi \alpha' F}{ 
g+2 \pi \alpha'F} \right) 
\frac{ |j| \tilde \omega^{h_j} }{ 1- \tilde \omega^{h_j} }
\right)
\right. \right. 
\nonumber \\
&~&
\left( \frac{1}{ \delta_{kj} -
|z|^2 F(k+1,j+1;2;|z|^2)  \left( \frac{g - 2\pi \alpha' F}{g+2 \pi \alpha'F} \right)^2
 \frac{ |k| 
\omega^{h_k} }{ 1- \omega^{h_k} }
|z|^2 F(k+1,l+1;2;|z|^2) } \right)
\nonumber \\
&~& \left. \left. \left( \delta_{0l}  - |z|^2 F(1,l+1;2;|z|^2) 
\left( \frac{g - 2\pi \alpha' F}{g+2 \pi \alpha'F} 
\right)
\frac{ |l| \tilde \omega^{h_l} }{ 1- \tilde \omega^{h_l} } \right) -1  \right] \right\}
\nonumber \\
&~& + O(g_o^2)
\eeqn
This is a generalization of the Born Infeld action taking into account higher loop stringy corrections,
specifically including contributions from euler number $\chi=3$ and including the contributions from
non-orientable surfaces such as the Mobius strip.
The construction presented in this work can be generalized without much effort to higher 
genus.	
It quickly becomes apparent that the simplifications obtained by the method of encoding the Greens
function in the boundary state are  overwhelmed by the increase in the parameters associted with the
various boundary states.

\section{Acknowledgements}

The author wishes to thank G. Semenoff, K. Buckley, P. DeBoer, and T. Davis
 for their helpful discussions on 
this work.
This work was supported in part by NSERC of Canada.

\bibliography{boundarys}

\end{document}